\documentstyle[manuscript,aps]{revtex}
\tightenlines
\begin{document}

\title{The exchange energy in the local Airy gas approximation}

\author{L. Vitos, B. Johansson}

\address{Condensed Matter Theory Group, Physics Department, \\
     Uppsala University, S-75121 Uppsala, Sweden}

\author{J. Koll\'ar}

\address{Research Institute for Solid State Physics, \\
H-1525 Budapest, P.O.Box 49, Hungary}

\author{H. L. Skriver}

\address{Center for Atomic-scale Materials Physics
and Department of Physics,\\
Technical University of Denmark, DK-2800 Lyngby, Denmark}

\date{4 May 2000}

\maketitle

\begin{abstract}
The Airy gas model of the edge electron gas is used to construct an
exchange-energy functional which is an alternative to those obtained
in the local density and generalized gradient approximations. Test
calculations for rare gas atoms, molecules, solids and surfaces show
that the Airy gas functional performs better than the local density
approximation in all cases and better than the generalized gradient
approximation for solids and surfaces.

\end{abstract}
\vspace{10mm}

\narrowtext
\pagebreak

Since the pioneering papers on density functional theory (DFT) 
\cite{hohenberg64,kohn65} there has been a constant search for 
exchange-correlation functionals of chemical accuracy. This includes the 
works on the generalized gradient approximation (GGA) 
\cite{perdew86,perdew91,perdew96,zhang98,perdew98} which are dedicated 
efforts to construct local functionals for inhomogeneous systems ranging 
from atoms to solids based on the uniform electron gas, i.e., the local 
density approximation (LDA), and density gradient corrections, as well as the
development of a number gradient level, semiempirical functionals 
\cite{lee88,becke88,hamprecht98,hammer99}. The GGA functionals have had a 
considerable impact upon the fields of quantum chemistry and solid state 
physics because they reduce the LDA overbinding and generally improve the 
calculated properties, relative to experiments, of molecules 
\cite{perdew96,perdew99,kurth99,becke86p} and bulk solids 
\cite{perdew92p,vitos97,ozolins93,kollar97,soderlind96,cho96}. However, they 
perform less well for the bulk properties of late transition metals and 
semiconductors \cite{khein95,corso96,filippi94}, and the underestimate of 
the exchange energies of surfaces \cite{perdew92p} as well as the 
overestimate of the dissociation energies of the multiply bonded molecules 
\cite{perdew96,kurth99} indicate the necessity to go beyond the
gradient level approximations and develop functionals that depend upon
other inhomogeneity parameters, e.g., higher derivatives of the charge
density or the Kohn-Sham kinetic energy density. One step in this direction
is the meta generalized gradient approximation (meta-GGA) of Perdew, Kurth, 
Zupan, and Blaha \cite{perdew99} which proves highly promising for both 
finite and extended systems \cite{kurth99}.

In the present work we introduce and apply a new gradient level exchange 
energy functional based on the concept of the edge electron gas \cite{kohn98}.
Besides the formal interest in the development of density based, orbital 
independent functionals there are several reasons why in applications of DFT 
the focus is on the approximate, local exchange-correlation schemes. Within 
the Kohn-Sham approach to DFT the Kohn-Sham exchange energy may be determined
exactly and as demostrated recently \cite{stadele97,gorling99,ivanov99} so 
may the corresponding local exchange potential. However, the exact Kohn-Sham 
exchange formalism is non-local and orbital-based, i.e. both the exchange 
energy and potential are highly complicated non-local functionals of the 
Kohn-Sham orbitals. In consequence, the application of exact exchange is 
computationally demanding. Furthermore, when exchange is treated exactly the 
error cancellations between the exchange and correlation energies on which 
all approximate schemes depend are lost owing to the poor description of 
correlation effects and, as a result, the total energies worsen 
\cite{perdew92p,gorling99}. For these reasons the exact Kohn-Sham exchange 
energy has only been used in practice in connection with semi-empirical, 
hybrid approximations \cite{becke}.

The concept of the edge electron gas was put forward by Kohn and Mattsson
\cite{kohn98} as an appropriate basis for the treatment of systems with edge
surface outside of which all Kohn-Sham orbitals decay exponentially. Its 
simplest realization, the Airy gas model, is based on the linear potential 
approximation and may serve as the starting point for the construction of 
functionals which are alternative to the GGA. The Airy gas model has recently 
been used to construct an explicit kinetic energy functional for 
inhomogeneous systems \cite{vitos99} which for atoms and surfaces has the 
accuracy of functionals based on a second order gradient expansion.

Here we have taken the exchange energy of the Airy gas model derived by Kohn
and Mattsson \cite{kohn98} and cast it in a form amenable to a simple,
accurate parametrization. The procedure may be viewed as local mapping of the
real system described by its density and scaled gradient onto the Airy gas
model and represents one possible solution to the joining of the interior
to the edge regions. The parametrized functional which we refer to as the 
local Airy gas (LAG) functional is tested in calculations of the exchange 
energies of rare gas atoms and of metallic surfaces within the jellium model 
where the exact results are known \cite{lang70}. In addition, we apply the 
LAG exchange functional in conjunction with the LDA for correlation 
\cite{perdew92} in calculations of the molecular binding energies and bulk 
properties of solids.

The present LAG exchange functional has a number of advantages over previous
GGA functionals: i) it explicitly includes the properties of the edge region
where much interesting physics occurs, ii) its accuracy may be systematically 
improved by including higher order expansions of the 
effective potential of the model system, and iii) the resulting 
exchange-energy functional is as simple and well-defined as that of the 
standard LDA. i.e., it has no adjustable parameters.

The starting point for the Airy gas exchange energy functional is the
potential

\begin{eqnarray}\label{eq:linearpot}
v_{eff}(z) = \left\{
\begin{array}{lll}
\infty\ &\mbox{for} & z \le -L \\
Fz      &\mbox{for} & -L < z < \infty
\end{array}\right.,
\end{eqnarray}
which is linear in $z$, independent of $x$ and $y$, and has a hard wall at
$-L$ far from the electronic edge at $z=0$. The slope of the effective 
potential $F = dv_{eff}/dz$ leads to a characteristic length scale

\begin{equation}\label{eq:l}
l\equiv \left(\frac{\hbar^2}{2mF}\right)^{1/3} ,
\end{equation}
and the electron and exchange-energy densities are then given by

\begin{equation}
n(z) = l^{-3}\;n(\zeta),
\end{equation}
and

\begin{equation}\label{eq:exdef}
\varepsilon_x(z) =-\frac{e^2}{2} l^{-4}\;\varepsilon_x(\zeta),
\end{equation}
where $\zeta=z/l$,

\begin{equation}
n(\zeta) = \;\frac{1}{2\pi}\int_0^{\infty} Ai^2(\zeta+\zeta')\ \zeta'd\zeta',
\end{equation}
and
\begin{eqnarray}\label{eq:exairy}
\varepsilon_x(\zeta)& = & \frac{1}{\pi}\int_{-\infty}^{\infty}
 \int_0^{\infty}\int_0^{\infty} Ai(\zeta+\epsilon)
 Ai(\zeta'+\epsilon) Ai(\zeta+\epsilon') Ai(\zeta'+\epsilon')
\nonumber \\
&\times& |\zeta'- \zeta|^{-3}
g(\sqrt{\epsilon}|\zeta'- \zeta|,\sqrt{\epsilon'}|\zeta'- \zeta|)
d\zeta' \; d\epsilon \; d\epsilon'.
\end{eqnarray}
A contour plot of the universal function $g(s,s')$ may be found in Ref.\
\cite{kohn98}. The exchange energy (\ref{eq:exdef}) may be written in the
form

\begin{equation}\label{eq:exF}
\varepsilon_x(z)=\varepsilon_x^{LDA}(z)F_x[s(z)],
\end{equation}
where $\varepsilon_x^{LDA}(z)$ is the exchange energy density of the uniform
electron gas. The enhancement function

\begin{equation}\label{eq:F}
F_x(\zeta) \equiv \frac{2}{3}\left (\frac{\pi}{3}\right )^{1/3}
\frac{\varepsilon_x(\zeta)}{n^{4/3}(\zeta)},
\end{equation}
is the unique function $(F_x(\zeta),s(\zeta))$ of the scaled gradient

\begin{equation}\label{eq:gradient}
s(\zeta) \equiv \frac{n'(\zeta)}{2(3\pi^2)^{1/3}n^{4/3}(\zeta)},
\end{equation}
plotted in Fig.\ \ref{fig1}. For comparison we also present results obtained
by the GGA of Perdew, Burke, and Ernzerhof (PBE) as defined in Ref.\
\cite{perdew96}, and the second order gradient expansion (GEA) \cite{gea}. It
follows from the figure that the exchange density (\ref{eq:exF}) in the low 
gradient limit of the Airy gas model reduces to $\varepsilon_x^{LDA}(z)$ as 
it should. In the large gradient limit 
$\varepsilon_x(\zeta) \approx -\frac{n(\zeta)}{2\zeta}$, Ref.\ \cite{kohn98},
and similar to the case of the kinetic energy density \cite{vitos99} we use
the properties of the Airy gas to find the following explicit asymptotic
expression

\begin{equation}\label{eq:large}
\varepsilon_x[n(z)] \approx -\frac{e^2}{2}\frac{n(z)}{2}\left[ n(z)
\frac{\partial^3 n(z)}{\partial z^3}
-\frac{\partial n(z)} {\partial z}\frac{\partial^2 n(z)}{\partial z^2}
\right] \left [
\frac{\partial n(z)} {\partial z}\right]^{-2},
\end{equation}
in terms of the density and its derivatives.

The density of the exchange energy per electron of the Airy gas is plotted as
a function of the distance $z$ from the electronic edge in Fig.\ \ref{fig2}.
It is seen that the large gradient expression (\ref{eq:large}) is accurate
for $z/l > -1.4$ corresponding to $s > 0.5$. It is also seen that neither the
LDA nor the PBE GGA \cite{perdew96} leads to the correct behaviour near and 
beyond the electronic edge at $z=0$.

The scaled gradient is conserved when going from the real electron gas to the
Airy gas model \cite{vitos99} and therefore the enhancement function $F_x(s)$
parametrized, for instance, in a modified Becke form \cite{becke86}

\begin{equation}\label{eq:exp}
F^{LAG}_x(s) = 1+\beta\frac{ s^{\alpha}}
{(1+\gamma s^{\alpha})^{\delta}},
\end{equation}
which includes the proper LDA limit, may be used to obtain the exchange 
energy density of the real electron gas from the local, scaled gradient 
$s[n(z)]$. For $\alpha= 2.626712, \beta=0.041106, \gamma= 0.092070$, and 
$\delta= 0.657946$ we find that the local deviation between the exact result 
(\ref{eq:F}) and the parametrized form (\ref{eq:exp}) integrated over 
the range $0<s<20$ is less than $0.3 \%$. We note that the present 
parametrization, being an overall fit, does not reduce to the GEA
\cite{perdew92p,dreizler90} in the low gradient limit. In contrast to the 
case of the kinetic energy \cite{vitos99} we have not been able to find an 
explicit, analytical expression for the exchange energy for small $s$ values, 
and to establish the behaviour numerically has not been attempted because the 
$s \rightarrow 0$ limit is reached only at $z\rightarrow -\infty$ as seen in 
Fig.\ \ref{fig2}. The exact behaviour of the LAG exchange functional at 
$s \rightarrow 0$ is therefore not know at present.

In the following we report the results of applying the LAG exchange 
functional to four test systems: i) rare gas atoms, ii) diatomic molecules, 
iii) jellium surfaces, and iv) solids. In all cases the total energy is 
calculated using self-consistent LDA densities. For molecules and solids 
the LAG exchange energy is combined with the LDA correlation energy 
\cite{perdew92}, since correlation effects has not been worked out in the 
Airy gas model. The motivation of this combination is given in terms of 
the enhancement function over the local exchange energy \cite{perdew96}, 
defined as $F_{xc}(s)\equiv\epsilon_{xc}[n]/\epsilon^{LDA}(n)$,
where $\epsilon_{xc}[n]$ denotes the exchange-correlation energy density.

Most of the currently applied approximate density functionals are based
on error cancellations between the exchange and correlation energies
\cite{kurth99,gorling99}. For physically interesting densities this
cancellation leads to $F_{xc}(s)$ with negligible slope up to $s\approx 1$.
Plots of the enhancement function over the local exchange energy 
for gradient level and meta-GGA approximations can be found in Refs. 
\cite{perdew96,kurth99}. In the present LAG exchange plus LDA correlation 
scheme this function becames 
$F^{LAG}_{xc}(s) = F^{LAG}_x(s) + \epsilon^{LDA}_c(n)/\epsilon^{LDA}_x(n)$,
where $\epsilon^{LDA}_c(n)$ is the correlation energy density of the uniform 
electron gas. Thus, the $F^{LAG}_{xc}(s)$ is determined only by the
LAG enhancement function (\ref{eq:exp}), which, for $s<1$, is a slowly 
increasing function of $s$. Therefore, we expect the present 
exchange-correlation scheme to preserve the excellent cancellation properties 
of the LDA and PBE GGA, and, at the same time, to bring the calculated
properties in closer agreement with experiment than conventional LDA.

For the rare gas atoms included in Table \ref{table1} the GEA, PBE, and LAG
functionals yield exchange energies which are, on the average, 6.4 \%,
8.5 \%, and 1.8 \%, respectively, larger than those obtained in the LDA.
The PBE values are in very good agreement with the exact Kohn-Sham results
\cite{kurth99,dreizler90}, which are given relative to the LDA energies
in the last column of the table. The LAG approximation represents only a 
minor improvement relative to the LDA total atomic exchange energies. 

The effect of the gradient correction to the LDA atomization energies for a 
few selected diatomic molecules is shown in Table \ref{table2} which also
includes the relative difference between the LDA results and experimental 
data \cite{kurth99}. Here, the LDA charge densities for the molecules have 
been generated using the full charge density (FCD) technique in conjunction
with the exact muffin-tin orbital method (EMTO) \cite{kollar00,andersen94,%
vitos00}. It is seen that the LAG approximation (i.e. LAG exchange and LDA
correlation energy) and PBE GGA have comparable 
accuracy: Both functionals reduce the LDA overbinding, and yield atomization 
energies which are, on the average, 16.8 \% (PBE) and 16.2 \% (LAG) smaller 
than the LDA values.

In Fig.\ \ref{fig3} we compare four exchange functionals applied to the
jellium model of metallic surfaces \cite{lang70}. The fact that for a given
$r_s$-value the exchange energies become increasingly negative in the order
LDA, LAG, GEA, and PBE is a simple consequence of the enhancement functions
shown in Fig.\ \ref{fig1} and in agreement with the observation that the GGA
significantly underestimate surface exchange energies \cite{perdew92p}.
We note that LAG approximation represents an improvement over both the LDA
and PBE and vary less with $r_s$ than either of the other two approximations.

As a final test of the LAG approximation we have calculated the atomic 
volumes and bulk moduli of several metals and semiconductors in their 
observed low temperature crystal structures by means of the FCD-EMTO method
\cite{kollar00,andersen94,vitos00}. The results for the equilibrium atomic 
radii are plotted in Fig.\ \ref{fig4}. For some selected metals and 
semiconductors, for which accurate LDA, PBE, and meta-GGA results have been 
published \cite{kurth99}, the atomic radii and bulk moduli are presented in 
Table \ref{table3}. The comparison of our LDA \cite{perdew92} atomic radii 
for the transition metal series with those obtained by the full-potential 
linear muffin-tin orbital \cite{ozolins93} and linear augmented plane wave 
\cite{khein95} methods using the same LDA gives mean deviations of 0.33 \%, 
0.43 \%, and 0.49 \% for the $3d, 4d$, and $5d$ series, respectively. For Li 
and Na the present LDA results agree within 0.07 \% with the full-potential 
values from Ref.\ \cite{perdew92p}. We therefore expect that the results of 
the present LAG and PBE calculations shown in Fig.\ \ref{fig4} will deviate
less than 0.5 \% from full-potential calculations. The mean deviations between 
the present atomic radii and bulk moduli listed in Table \ref{table3} and 
those of Ref.\ \cite{kurth99} obtained using the linear augmented plane wave 
method are 0.20 \% and 3.28 \% for the LDA and 0.27 \% and 3.26 \% for the 
PBE functionals.

The LDA atomic radii shown in Fig.\ \ref{fig4} deviate, on average, by
2.26 \% from the experimental values \cite{young91,radii0}, while those
calculated in the LAG model
and the PBE deviate by $0.83 \%$ and $0.91 \%$, respectively. Among the
energy functionals considered in Table \ref{table3} the LAG is found to
give the lowest mean deviations for both atomic radii and bulk moduli.
We note that for these solids the LAG approximation achieves the accuracy of
the recently developed meta-GGA \cite{perdew99,kurth99}.

We have used the Airy gas model of the edge electron gas that is equivalent
to the linear potential approximation to develop an exchange energy 
functional which may serve as an alternative to the functionals based on the 
generalized gradient appoximation, e.g., PBE GGA. Test calculations for 
finite and extended systems show that the LAG approximation is more accurate 
than the local density approximation in all cases. While the LAG results for 
atoms are very close to the LDA results and, hence inferior to the PBE GGA
results, its accuracy for the atomization energies of diatomic molecules is 
similar to that of the PBE GGA. In bulk systems the LAG results are, on 
average, closer to the experimental values than those obtained in the 
PBE GGA. These results are very satisfactory in view of the fact that the LAG
exchange functional is derived solely form the properties of the Airy gas,
and, hence, with no {\it a priory}\/ assumptions concerning the exchange 
enhancement factor. In this sense it is truely {\it ab initio} but for the 
correlation effects which needs to be worked out in the Airy gas model.

{\it Acknowledgments:} We gratefully acknowledge interesting and fruitful
discussions with Professor \ Walter Kohn. L.V. acknowledges the valuable
observations by Professor J. P. Perdew and Dr. \'A. Nagy.
L.V. and B.J. are grateful to the
Swedish Natural Science Research Council and the Swedish Foundation for
Strategic Research for financial support. Part of this work was supported
by the research project OTKA
23390 of the Hungarian Scientific Research Fund. The Center for Atomic-scale
Materials Physics is sponsored by the Danish National Research Foundation.

\begin{table}
\footnotesize
\caption[1]{The effect of GEA \cite{gea}, PBE \cite{perdew96}, and LAG 
gradient corrections (in percentage) on the LDA 
atomic exchange energies. All functionals are evaluated from the 
self-consistent LDA \cite{perdew92} Kohn-Sham densities. KS denotes the 
relative difference of the exact and LDA exchange energies from Ref.\ 
\cite{kurth99}.}
\normalsize
\begin{tabular}{lcccc}
ATOM & GEA &  PBE & LAG & KS\\
\hline
\hline
He & 13.9 & 15.0 & 4.2 & 16.0 \\
Ne &  6.7 & 9.4  & 1.9 & 9.7 \\
Ar &  5.1 & 7.7  & 1.4 & 8.3 \\
Kr &  3.5 & 5.5  & 0.9 & 5.9 \\
Xe &  2.9 & 4.7  & 0.7 & 5.0 \\
\end{tabular}
\label{table1}
\end{table}

\begin{table}
\footnotesize
\caption[2]{The effect of PBE \cite{perdew96} and LAG gradient corrections
(in percentage) on the LDA atomization energies for diatomic molecules.
Both functionals are evaluated from the self-consistent LDA \cite{perdew92}
Kohn-Sham densities generated by the FCD-EMTO 
\cite{kollar00,andersen94,vitos00}. Expt. denotes the relative difference of 
the experimental and LDA atomization energies from Ref.\ \cite{kurth99}.}
\normalsize
\begin{tabular}{lcccc}
MOLECULE & PBE & LAG & Expt.\\
\hline
\hline
Li$_2$& -22.6 & -17.8 &   2.1 \\
Be$_2$& -25.9 & -32.8 & -76.6 \\
CO    &  -9.8 &  -8.6 & -13.3 \\
N$_2$ & -12.2 & -10.7 & -14.5 \\
NO    & -14.2 & -11.2 & -23.0 \\
O$_2$ & -16.1 & -16.0 & -31.1 \\
\end{tabular}
\label{table2}
\end{table}

\begin{table}
\footnotesize
\caption[3]{Theoretical equilibrium atomic radii (in $Bohr$) and bulk moduli
(in $GPa$) for some selected solids. The present calculations have been
performed for crystallographic $\alpha$ phases using the FCD-EMTO method 
\cite{kollar00,andersen94,vitos00}. The results obtained by the meta-GGA 
of Perdew, Kurth, Zupan and Blaha (PKZB) and the experimental values are 
from Ref.\ \cite{kurth99}. The mean absolute values of the relative errors 
are shown in parenthesis.
}
\normalsize
\begin{tabular}{lcccccccccc}
  &S$_{LDA}$&S$_{PBE}$&S$_{LAG}$&S$_{PKZB}$&S$_{Expt.}$&B$_{LDA}$&B$_{PBE}$
  &B$_{LAG}$&B$_{PKZB}$&B$_{Expt.}$\\
\hline
\hline
Na    &3.769&3.916&3.927&4.019&3.936&  8.2&  7.6&  7.3&  7.0&  6.9 \\
Al    &2.947&2.989&2.977&2.966&2.991& 81.2& 75.2& 76.2& 90.5& 77.3 \\
Fe    &2.565&2.645&2.604&2.627&2.667&253  &178  &209  &198  &172   \\
Cu    &2.602&2.684&2.656&2.656&2.658&193  &137  &157  &154  &138   \\
Pd    &2.846&2.916&2.883&2.888&2.873&235  &184  &203  &181  &181   \\
W     &2.929&2.977&2.953&2.946&2.940&312  &292  &299  &311  &310   \\
Pt    &2.888&2.943&2.916&2.908&2.892&304  &244  &268  &267  &283   \\
Au    &2.998&3.081&3.043&3.041&2.997&194  &134  &156  &153  &172   \\
Si    &3.163&3.198&3.189&3.200&3.182&100  & 92.8& 94.0& 93.6& 98.8 \\
Ge    &3.303&3.384&3.354&3.349&3.318& 71.6& 61.2& 64.0& 64.6& 76.8 \\
GaAs  &3.296&3.375&3.346&3.347&3.312& 73.0& 62.0& 72.1& 65.1& 74.8 \\
NaCl  &3.202&3.346&3.337&3.284&3.306& 32.9& 23.0& 21.7& 28.1& 24.5 \\
    &(1.48 \%)&(1.28 \%)&(0.80\%)&(0.88 \%)&&($17.2 \%$)&($9.2 \%$)&($9.1 \%$)
    &(9.3 \%)&  \\
\end{tabular}
\label{table3}
\end{table}

\begin{figure}
%\centerline{\psfig{figure=fig1.ps,height=2.6in}}
\caption[1]{The exchange energy enhancement function (\ref{eq:F}) and the
parametrized form (\ref{eq:exp}), the latter indicated by dots, of the local
Airy gas (LAG) compared to those of the local density approximation (LDA),
the the generalized gradient approximation (PBE) as defined in Ref.\ 
\cite{perdew96}, and the second order gradient expansion (GEA) \cite{gea}.
}
\label{fig1}
\end{figure}

\begin{figure}
%\centerline{\psfig{figure=fig2.ps,height=3.0in}}
\caption[2]{The exchange energy per electron, $-\varepsilon_x(z)/n(z)$, of
the Airy gas, obtained from the enhancement factors shown in Fig.\
\protect\ref{fig1}, plotted as a function of the distance from the electronic
edge and compared to the exact result obtained from (\ref{eq:exdef}) and
(\ref{eq:exairy}) and the explicit large gradient limit (\ref{eq:large}).
Energy in units of $-(e^2/2)l^{-1}$ and distance in units of $l$ defined in
(\ref{eq:l}). The scaled gradient $s$ is also shown.
}
\label{fig2}
\end{figure}

\begin{figure}
%\centerline{\psfig{figure=fig3.ps,height=2.0in}}
\caption[3]{The exchange energy of the LDA \cite{perdew92}, GEA \cite{gea},
PBE \cite{perdew96}, and LAG approximations obtained from the self-consistent
Kohn-Sham densities relative to the exact exchange energy \cite{estim} in the 
jellium surface model for a range of density parameters 
$r_s = (3/4\pi n)^{1/3}$.}
\label{fig3}
\end{figure}

\begin{figure}
%\centerline{\psfig{figure=fig4.ps,height=5.0in}}
\caption[4]{Relative deviations of the calculated and experimental
\cite{young91} equilibrium atomic radii for the alkali, alkaline earth,
$3d, 4d$, and $5d$ transition metals using LDA \cite{perdew92}, PBE
\cite{perdew96} and LAG energy functionals.
The numbers in parenthesis are the average deviations.
}
\label{fig4}
\end{figure}


\begin{references}
\bibitem{gea} Here we use the second order gradient expansion enhancement 
factor, $1 + C_xs^2$ with $C_x =0.1234$ which according to Ref.\ 
\protect\cite{dreizler90}, pg.\ 87, should be preferred over the original
$C_x=0.0864$. In fact, we find that the original value performes better for 
both the Airy gas and jellium models.
\end{references}
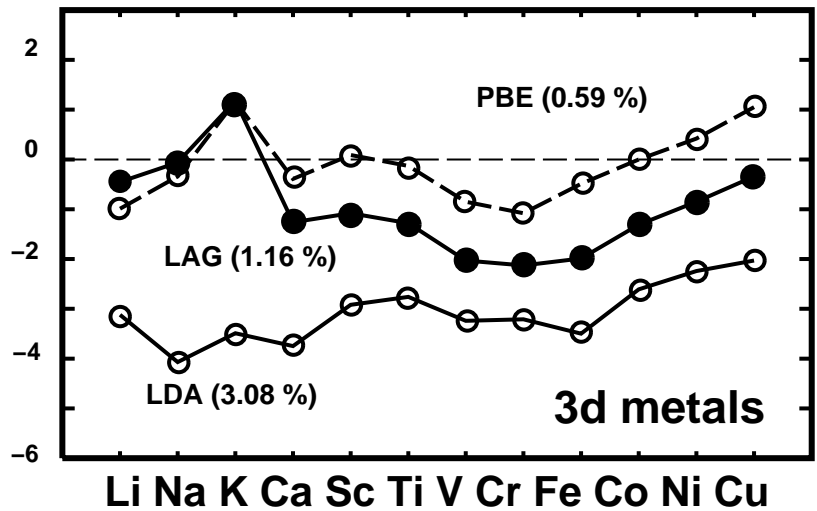
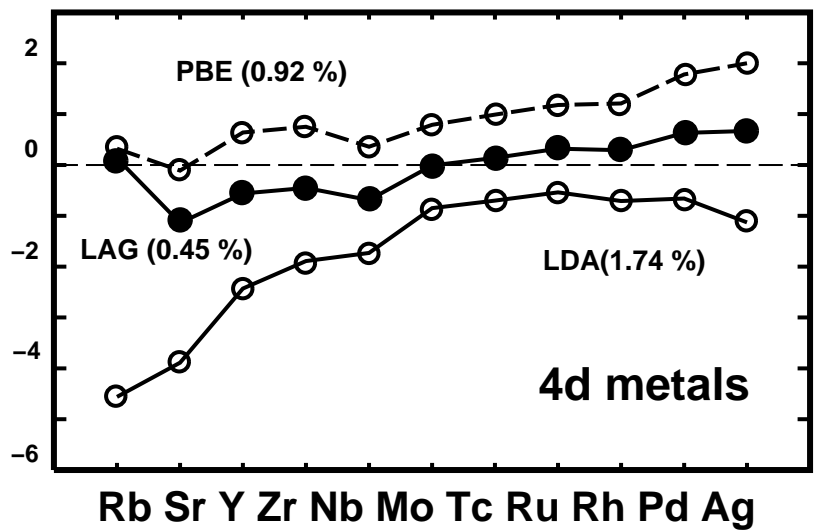
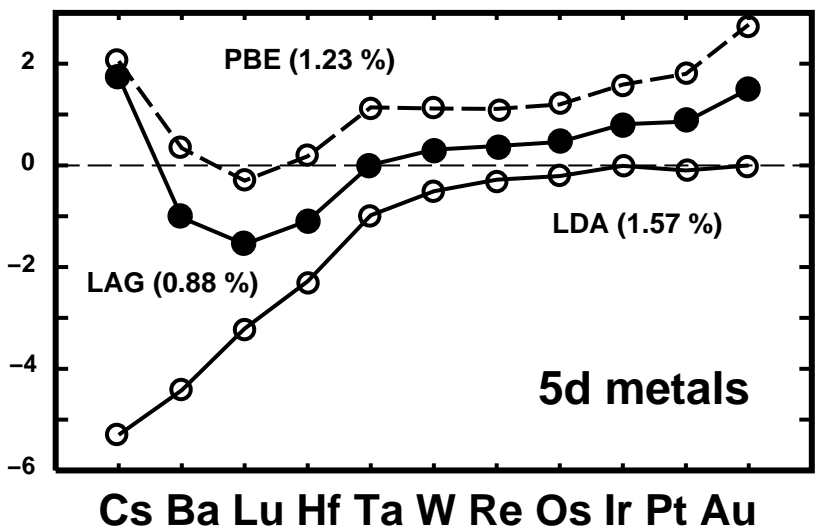
\end{document}